\documentclass[pra,aps,twocolumn,tightlines,showpacs]{revtex4}
\usepackage{amsmath}
\usepackage{epsfig}

\begin{document}

\title{RF-Photonic Frequency Stability Gear Box}

\author{A. B. Matsko$^*$, A. A. Savchenkov, V. S. Ilchenko, D. Seidel, and L. Maleki}

\affiliation{OEwaves Inc., 465 N. Halstead St. Ste. 140, Pasadena, CA 91107}

\begin{abstract}
An optical technique based on stability transfer among modes of a monolithic optical microresonator is proposed for  long therm frequency stabilization of a radiofrequency (RF) oscillator. We show that locking two resonator modes, characterized with dissimilar  sensitivity in responding to an applied forcing function, to a master RF oscillator allows enhancing the long term stability of a slave RF oscillator locked to two resonator modes having nearly identical sensitivity. For instance, the stability of a $10$~MHz master oscillator characterized with Allan deviation of $10^{-7}$ at $10^4$~s can be increased and transferred to a slave oscillator with identical stability performance, so that the resultant Allan deviation of the slave oscillator becomes equal to $10^{-13}$ at $10^4$~s. The method does not require absolute frequency references to achieve such a performance.
\end{abstract}

\maketitle

Stable oscillators are required in every branch of modern science and technology, so the demand for creation of novel devices with higher performance and smaller size, weight, and power consumption is continuously growing. On the other hand, the stability of simple free running electronic RF systems is limited both technically and fundamentally. The increasing complexity of  stabilization circuitry as well as usage of external reference units, e.g. atomic cells, is typically required to improve the performance of conventional electronic oscillators. In this Letter we disclose an approach for stabilization of an RF oscillator using an RF photonic transducer consisting of an optical microresonator possessing at least two families of modes with different sensitivity to an applied forcing function.

An optical resonator can be used as a transducer between optical and radio frequencies \cite{bay72prl}. Nearly three decades ago it was realized that locking the free spectral range (FSR) of a Fabry-Perot resonator to a radio frequency oscillator, and simultaneous locking a laser frequency to an optical mode of the resonator, results in the stabilization of the laser frequency \cite{devoe84pra}, since the optical frequency $\omega_0$ and the FSR frequency $\omega_{FSR}$ are related as
\begin{equation}
\omega_0= l\ \omega_{FSR}+\Delta \omega,
\end{equation}
where $l$ is an integer number, and $\Delta \omega$ is the frequency shift arising due to the dispersive phase shift corrections within the resonator. In other words, the modes of the resonator serve as a bridge between RF and the optical frequencies \cite{ye03aamop}.

The approach was devised  for an accurate transduction of a known RF frequency, precisely defined by an RF clock, to the optical frequency domain. Implementation of the method was problematic since it required an accurate knowledge of the resonator dispersion frequency dependent parameter $\Delta \omega$, which results from the dispersion associated with  mirror coatings. It became clear that the FSR of the resonator cannot be used as a constant frequency marker, and the resonator frequency bridges were replaced with optical frequency combs \cite{hall04ch} allowing transfer of frequency and phase information from the RF to the optical frequency domain.

While optical resonators are not particularly suitable for absolute frequency metrology, they are useful for oscillator stabilization. Locking to a reference cavity is widely utilized for stabilizion of laser frequency \cite{drever83apb,salomon88josab} and RF oscillators \cite{woode96tuffc}. It was demonstrated \cite{maleki11ifcs} that locking a mode of an optical monolithic microresonator to an optical frequency reference results in suppression of the long term frequency drift of the FSR of the resonator, as well as stabilization of the RF hyper-parametric oscillator based on the resonator.

An absolute frequency reference is not a prerequisite for the long term frequency stabilization of an oscillator. Application of optical dual-mode stabilization technique, initially developed for stabilization of RF quartz oscillators \cite{walls95uffc}, results in excellent long term stabilization of the spectrum of an optical microresonator, and an oscillator locked to this resonator \cite{savchenkov07josab,matsko11pra,strekalov11arch}. In this Letter we introduce another application of the optical dual mode technique, and propose to use it for long term stabilization of an RF oscillator. The approch involves an optical microresonator that enhances the efficiency of the electronic feedback used to stabilize the oscillator.

The microresonator-based RF photonic transducer uses two families of optical modes having significantly different susceptibility to external conditions such as mechanical pressure, voltage, temperature, etc. We select two arbitrary modes, each of which belongs to one the two mode families, and lock the frequency difference between those modes to a master RF oscillator. The locking can be implemented in various ways. For instance, let us assume that light emitted by a continuous wave (cw) laser is modulated with the RF signal. The carrier of the modulated light is locked to the center of one of the optical modes using, e.g., the Pound-Drever-Hall technique \cite{drever83apb}. An external parameter(s) is (are) adjusted via an electronic feedback so that the second selected optical mode has a  frequency equal to the frequency of the modulation sideband. In this way the stability of the microresonator spectrum becomes strongly correlated with the stability of the master RF oscillator, i.e. the spectrum of the resonator becomes stable.

Let us estimate in an explicit way the degree of the stability realized by this kind of locking.  We assume that the laser is tunable so its frequency ($\omega_0$) follows the corresponding optical mode. Locking of  frequencies of the modulation sideband and the other selected resonator mode ($\omega_1$) is realized via feedback to uncorrelated environmental (or applied) parameters $q_j$ changing the frequency of the both resonator modes. The drift of the optical frequencies can be expressed as
\begin{equation} \label{deltaopt}
\Delta \omega_{i}= \sum \limits_j \alpha_{i,j} \Delta q_j + \delta \omega_{i},
\end{equation}
where $i=0,1$, $\alpha_{i,j}$ are the scaling parameters, $\Delta q_j$ are the residual drifts of the corresponding environmental or applied parameters. Here we also took into account that the frequencies of carrier  and  modulation sideband  depend on the imperfection of the electronic locking, characterized via unknown detunings between the frequencies of the optical harmonics and the corresponding resonator modes ( $\delta \omega_{i}$). By definition, the frequency difference between the carrier and the modulation sideband is given by
\begin{equation} \label{deltarf}
\Delta \omega_{0} - \Delta \omega_{1}= \delta \omega_{RF},
\end{equation}
where $\delta \omega_{RF}$ is the residual frequency drift of the master oscillator. The basic goal of the locking procedure is to minimize either $\langle (\Delta \omega_{0})^2 \rangle $ or $\langle (\Delta \omega_{1})^2 \rangle$ quadratic deviation (\ref{deltarf}).

This problem has an explicit solution if only one environmental (or applied) parameter, e.g. the temperature of the resonator $T$ ($q_1\equiv T$), is important. This is the most practical case since the drift of the ambient temperature usually results in frequency drift of  the oscillator locked to the microresonator. We infer from Eq.~(\ref{deltarf}) that the residual temperature drift is given by the locking circuit $\Delta T=(\delta \omega_{RF}-\delta \omega_{0}+\delta \omega_{1})/(\alpha_{0,T}-\alpha_{1,T})$, and the stability of the optical harmonics is
\begin{equation} \label{deltaopt1}
\Delta \omega_{i}= \frac{\alpha_{i,T}}{\alpha_{0,T}-\alpha_{1,T}} (\delta \omega_{RF}-\delta \omega_{0}+\delta \omega_{1})  + \delta \omega_{i}.
\end{equation}
Ultimately, if $\alpha_{0,T}$ and $\alpha_{1,T}$ are significantly different, the stability of the optical harmonics is given by the stability of the RF master oscillator, $\langle (\Delta \omega_{i})^2 \rangle \sim \langle (\delta \omega_{RF})^2 \rangle$. Therefore, the technique allows transferring the stability of an RF oscillator to the optical domain in such a way that the {\em relative stability of the laser locked to the resonator becomes much larger compared with the relative stability of the RF master oscillator}.

\begin{figure}[ht]
 \center{
\epsfig{file=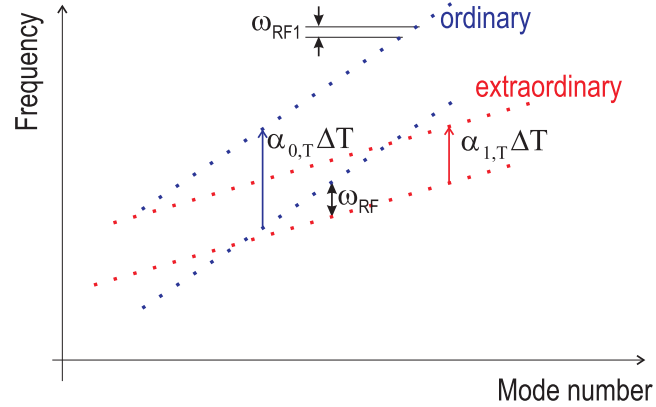, width=8.5cm, angle=0} }
\caption{\label{fig1}
Illustration of the thermal shift of the spectrum of a MgF$_2$ WGM microresonator. If the temperature of the resonator changes by $\Delta T$, the ordinarily polarized modes shift by $\alpha_{0,T} \Delta T$ in frequency, while extraordinarily polarized modes shift by $\alpha_{1,T} \Delta T$. To stabilize the frequency of the resonator spectrum, we propose to lock the frequency difference between one mode belonging to the family of ordinarily polarized modes, and the other mode belonging to the family of ordinarily polarized modes, to an RF master oscillator having frequency $\omega_{RF}$. This kind of locking not only stabilizes the entire optical spectrum, but also enhances the relative stability of the frequency difference between any two optical modes belonging to the same mode family (e.g. $\omega_{RF1}$, as shown in the picture) beyond the stability of the master oscillator.
}
 \end{figure}

It is useful to estimate the efficiency of the method. Let us consider a freely suspended MgF$_2$ WGM microresonator \cite{savchenkov07josab} and assume that $\omega_{0}$ ($\omega_{1}$) is the frequency of its ordinarily (extraordinarily) polarized mode, and $\delta \omega_{0}$ and $\delta \omega_{1}$ are negligible, so that
\begin{eqnarray} \label{o}
\alpha_{0,T} = -\frac{1}{n_o} \frac{\partial n_o}{ \partial T} -\frac{1}{R} \frac{\partial R}{ \partial T}, \\
\alpha_{1,T} = -\frac{1}{n_e} \frac{\partial n_e}{ \partial T} -\frac{1}{R} \frac{\partial R}{ \partial T}, \label{e}
\end{eqnarray}
where $n_e$ and $n_o$ are extraordinary and ordinary refractive indices of the material, and $R$ is the radius of the resonator (see Fig.~\ref{fig1}). The requirement of free suspension is needed to exclude from the consideration the thermally-dependent strain of the resonator. Taking into account $(\partial n_o/\partial t)/n_o= 0.6$~ppm/K,  $(\partial n_e/\partial t)/n_e= 0.25$~ppm/K, and $(\partial R/\partial T)/R=9$~ppm/K we find $\langle (\Delta \omega_{0})^2 \rangle^{1/2} \sim 27\langle (\delta \omega_{RF})^2 \rangle^{1/2}$ if the frequency of the light corresponds to 1.5~$\mu$m wavelength. The long term drift of the laser becomes only an order of magnitude larger compared with the long term drift of the master RF oscillator.
%
%

The approach is similar to the technique described in \cite{devoe84pra}, since the resonator operates as a transducer of the stability of an RF master oscillator to the optical frequency domain. However, the transduction efficiency of the approach proposed here is orders of magnitude higher. In fact in the approach described in \cite{devoe84pra} the resonator FSR was locked to an RF oscillator. In this case the frequency drift of the optical modes exceeds $(\omega_0/\omega_{RF})\delta \omega_{RF}$, which is a rather large value compared with $27\delta \omega_{RF}$.

The proposed locking technique is efficient for suppression of the frequency drift because of a single parameter $q_i$. If two independent drifting parameters are present (e.g. $T$ and $q$), and temperature $T$ is the parameter used in the feedback loop, in accordance with Eqs.~(\ref{deltaopt}) and (\ref{deltaopt1}), we find
\begin{equation}
\Delta \omega_{0}= \frac{\alpha_{0,T}}{\alpha_{0,T}-\alpha_{1,T}} \delta \omega_{RF}+ \frac{\alpha_{0,T}\alpha_{1,q}-\alpha_{1,T}\alpha_{0,q}}{\alpha_{0,T}-\alpha_{1,T}} \Delta q,
\end{equation}
where for the sake of simplicity we have neglected  $\delta \omega_{0}$ and $\delta \omega_{1}$. The technique helps to suppress the drift of parameter $q$ only if it influences the resonator modes involved in the locking process such that $\alpha_{0,T}\alpha_{1,q} \approx \alpha_{1,T}\alpha_{0,q}$. Therefore, to cancel the drift of $q$ one needs to create another locking loop. For instance, if one manages to reach $\Delta T=\xi (\delta \omega_{RF}-\delta \omega_{0}+\delta \omega_{1})/(\alpha_{0,T}-\alpha_{1,T})$ and $\Delta T=(1-\xi) (\delta \omega_{RF}-\delta \omega_{0}+\delta \omega_{1})/(\alpha_{0,q}-\alpha_{1,q})$ using two different electronic feedback loops, where $1 > \xi > 0$, the complete stabilization of the optical frequency drift will be achieved again. The only necessary condition is the asymmetry of the resonator response to the environmental or applied parameters  $\alpha_{0,i} \ne \alpha_{1,i}$

We have shown that it is possible to transfer the absolute stability of a master RF oscillator to a mode of an optical microresonator. Developing this idea further, we use the properties of an optical resonator to suggest the relative drift of the FSR and the optical frequencies belonging to the same mode family, are the same \cite{bay72prl,devoe84pra}
\begin{equation}
\frac{\Delta \omega_{FSR}}{\omega_{FSR}} = \frac{\Delta \omega_{0}}{\omega_{0}}.
\end{equation}
Therefore, using the stabilization procedure described above we are able to stabilize the resonator FSR such that
\begin{equation} \label{fsrdrift}
\frac{\Delta \omega_{FSR}}{\omega_{FSR}}= \frac{\alpha_{0,T}}{\alpha_{0,T}-\alpha_{1,T}} \frac{\delta \omega_{RF}}{\omega_{RF}} \frac{\omega_{RF}}{\omega_0}.
\end{equation}
In other words, {\em the long term stability of the FSR exceeds the stability of the master RF oscillator if the proposed stabilization procedure is used.}

In the particular case when the ordinarily and extraordinarily polarized resonator modes are used for locking, a stronger condition compared with (\ref{fsrdrift}) can be derived. For any mode pair having the same polarization and frequencies $\omega_1$ and $\omega_2$ the relative drift of the frequency difference $\omega_1-\omega_2=\omega_{RF1}$ is the same as the relative drift of the optical frequency. Selecting $\omega_{RF1}=\omega_{RF}$ we conclude that the optical resonator stabilized by locking to a master RF oscillator has two optical modes with relative long term frequency stability much better than the stability of the master oscillator:
\begin{equation} \label{RFdrift}
\frac{\Delta \omega_{RF1}}{\omega_{RF1}}= \frac{\alpha_{0,T}}{\alpha_{0,T}-\alpha_{1,T}}  \frac{\omega_{RF}}{\omega_0} \frac{\delta \omega_{RF}}{\omega_{RF}}.
\end{equation}
Those two stable modes can be used for stabilization of either a slave RF oscillator or the same master oscillator. In the case of slave oscillator one just need to lock it to the corresponding pair of optical modes. In the case of the stabilization of the master RF oscillator, the optical resonator should be inserted into the electronic feedback loop. Proper time constants and gains of the loops should be selected to achieve the optimal level of stabilization.

Let us consider a $10$~MHz master oscillator characterized with Allan deviation of $10^{-7}$ at $10^4$~s. In accordance with Eq.~(\ref{RFdrift}) the optical resonator made out MgF$_2$ allows achieving relative stability for two ordinary optical modes separated by 10~MHz at the level characterized by Allan deviation of $1.4 \times 10^{-13}$ at $10^4$~s. This long term stability can be transferred to a slave RF oscillator locked to the optical modes with an opto-electronic feedback loop.

To conclude, we have shown that an optical resonator can be used as a frequency stability transducer for RF oscillators. Creation of the suggested opto-electronic circuits will allow  ultra-stable compact RF oscillators withoutthe need for absolute frequency references. We believe this technique can significantly improve applications where small and efficient highly stable oscillators are required.

\end{document}